\documentclass[floatfix,pra,twocolumn,amsmath,amssymb,letterpaper,groupaddresses,superscriptaddress]{revtex4}
\usepackage{graphicx}
\usepackage{hyperref}
\usepackage{xcolor}
\usepackage{mathtools}
\usepackage{times}
\usepackage{latexsym}
\usepackage{graphicx}
\usepackage{amsmath,amssymb}
\usepackage{verbatim,times,bbm}
\usepackage{placeins}
\usepackage{color}
\usepackage[english]{babel}
\usepackage[T1]{fontenc}
\usepackage{amsmath, amsthm, amssymb, mathrsfs}
\usepackage{epstopdf}
\usepackage{subfigure}
\usepackage{appendix}
\usepackage{pdfpages}
\usepackage{multirow}
\usepackage{placeins}
\usepackage{float}

\usepackage{times}
\usepackage{latexsym}
\usepackage{graphicx}
\usepackage{verbatim,times,bbm}
\usepackage{color}
\usepackage{appendix}
\usepackage{graphicx}% Include figure files
\usepackage{dcolumn}% Align table columns on decimal point
\usepackage{breqn}
\usepackage{comment}
\usepackage{subfigure}
\usepackage{tikz}
\usepackage{epsfig}
\usepackage{color}
\usepackage{bm}% bold math
\usepackage{placeins}
\usepackage{float}
\usepackage{color}

\newcommand\beq{\begin{equation}}
\newcommand\eeq{\end{equation}}
\newcommand\bea{\begin{eqnarray}}
\newcommand\eea{\end{eqnarray}}

\newcommand{\ceil}[1]{\left\lceil #1 \right\rceil}
\newcolumntype{?}{!{\vrule width 1.3pt}}
\begin{document}
%---------------------------------------------------------------------
\title{Breaking RSA Security With A  Low Noise D-Wave 2000Q Quantum Annealer:\\ Computational Times, Limitations And Prospects}

\author{ Riccardo Mengoni}\email{r.mengoni@cineca.it}
\author{ Daniele Ottaviani}\email{d.ottaviani@cineca.it}
\affiliation{CINECA$, $ Casalecchio di Reno$ ,  $ Bologna$, $  Italy} 

\author{ Paolino Iorio}\email{paolino.iorio@eni.com}
\affiliation{ENI S$.$p$.$A$.$$ , $ San Donato Milanese$ , $ Milano$ , $ Italy}

%---------------------------------------------------------------------

%---------------------------------------------------------------------
\maketitle

%---------------------------------------------------------------------
\section*{Abstract }
The RSA cryptosystem  could be easily broken with   large scale general purpose quantum computers     running Shor's factorization  algorithm. Being such devices still in their infancy, a  quantum annealing approach to   integer factorization has recently gained attention. In this work, we  analysed  the most promising  strategies  for RSA hacking  via  quantum annealing with an extensive study of  the \textit{low noise D-Wave 2000Q} computational times,  current hardware limitations and  challenges for  future developments.

\section{Introduction }
 \label{ProblemIntro}
In 1977,  Ronald Rivest, Adi Shamir and Leonard Adleman proposed an algorithm  for securing data, better  known today with the acronym of RSA,  which became     the most famous and widely used algorithm in security \cite{RSA}.

The core idea behind RSA comes from a well known  problem in number theory, the \textit{Prime Factorization (PF) problem}, stating the following.  
\\\\
\textit{Consider an integer $ N $ which is the product of two unknown prime numbers. The task is to find the primes $ p $ and $ q $ such that
\begin{equation*}
N=p\times q
\end{equation*}}
This seemingly easy task turns out to be hard in practice, in fact it does not exist a deterministic algorithm that finds the prime factors in polynomial time \cite{Krantz,Arora}.
The best classical algorithm \cite{Lenstra} is able to factor  an integer $ N $ in a time that is
\begin{equation}
{\displaystyle O\!\left(e^{1.9(\log N)^{1/3}(\log \log N)^{2/3}}\right)}.
\end{equation}
Hence the complexity class associated to the PF problem is   NP, despite not considered to be  NP-complete \cite{Goldreich}.
 
In 1994, a  groundbreaking result was published by the mathematician Peter Shor \cite{Shor}.  He proved that a general purpose quantum computer could solve the integer factorization problem exponentially faster than the best classical algorithm. 
Shor's quantum algorithm cleverly employs the  quantum Fourier transform subroutine to factor an integer  $ N $ in a number of steps of  order
\begin{equation}
{\displaystyle O\!\left((\log N)^{2}(\log \log N)(\log \log \log N)\right)} .
\end{equation}
This means that, in order to break RSA-2048, which involves the factorization of a 2048 binary digits number, the best classical algorithm would require around $  10^{34} $ operations, while  a quantum computer only needs $ 10^8 $ quantum operations.
	
However, it is largely believed that a physical implementation of a fault tolerant general purpose quantum computer able to run Shor's algorithm on large integers and break RSA-2048 is still decades away \cite{gidney2019factor}. 

On the other side, a large effort has been put forward in the development of  annealers i.e. specific purpose quantum  devices  able to perform optimization and sampling tasks.
In the last few years, multiple implementations of  Shor-like algorithm running on quantum annealers have been proposed and tested \cite{Schaller,Dattani,Jiang}. Still, a precise report of the running times required by the annealing hardware  to factor integers is currently lacking  in the literature.

In this work, we aim at analysing  the most promising  approaches to integer factorization via  quantum annealing with an extensive investigation of  the \textit{low noise D-Wave 2000Q} computational times,  current limitations of the hardware,  opportunities and challenges for the future.

This paper is organized as follows: in the first Section, the technique of Quantum Annealing is introduced, as well as the properties of the latest D-Wave quantum annealing device. In Section \ref{QUBO}, several formulations of the  factorization problem are presented. Finally, in Section \ref{PrimeDWave}, our approach for solving    PF  with the D-Wave 2000Q quantum annealer is explained  and analysed in detail.

\section{Quantum Annealing}
 \label{QA}
 %\subsection{Quantum Annealing} 
 Several  algorithms have been proposed   for   finding the global minimum of a given objective function with several local minima.
 
 An example is Simulated Annealing (SA), where thermal energy is used  to escape local minima and  a   cooling process  leads the system towards  low energy states \cite{SA}.
 The transition  probability between minima  depends  on the height $ h $ of the potential well   that separates them, $ e^{-\frac{h}{k_B T}} $. Hence  SA is  likely to get stuck  in the presence of very high barriers.
%Adiabatic quantum computation (AQC) is a model of quantum computation which relies on the adiabatic theorem in order to perform calculations. It is closely related to

 Similarly, the   meta-heuristic technique known as Quantum Annealing (QA) searches for  the global minimum of an objective function by  exploiting quantum tunnelling in the  analysis of the    candidate solutions space \cite{QA}. %Like in  simulated annealing,  whose temperature parameter determines the probability of jumping between local minima, in QA the  tunnelling field strength is responsible for the transition to better minima. 

The advantage is that  the tunnelling probability depends both on the height $ h $ and the width $ w $ of the potential barriers, $ e^{-\frac{w\sqrt{h}}{\Gamma}} $, where $ \Gamma $ is the transverse field strength. This  gives  QA the ability to easily move in an energy landscape where local minima  are separated  by tall barriers, provided that they are narrow enough \cite{PhysRevX.6.031015}.

%\subsection{Adiabatic Quantum Computation}
QA is strictly related to Adiabatic Quantum Computing (AQC), a   quantum computation scheme where a  system is initialized to the ground state of a  simple  Hamiltonian and then gradually turned to reach a desired problem Hamiltonian \cite{McGeoch}.  
AQC is based on the   adiabatic theorem stating that, if  the quantum evolution is slow enough,   the system  remains in  its ground state throughout the whole process.

However, while  AQC assumes a unitary and adiabatic quantum evolution,  QA allows  fast evolution   exceeding the adiabatic regime. For this reason, plus the fact that   temperature  is often few mK above  absolute zero, QA is  not guaranteed to  end up the annealing process in the system ground state.

Formally speaking, the  time-dependent Hamiltonian describing    QA is the following
\begin{align}
H(t) =  A(t) H_0 + B(t) H_{\rm P}\;,
\label{eq:H}
\end{align}
where $ H_0 $ and   $H_{\rm P}$ are  respectively the initial and problem  Hamiltonian.  The annealing schedule is controlled by the functions $A(t)$ and $B(t)$, defined in the   interval $t \in [0,T_{\rm QA}]$, where $ T_{\rm QA} $ is the total annealing time.
 
\begin{figure}[htbp]
	\includegraphics[scale=0.6]{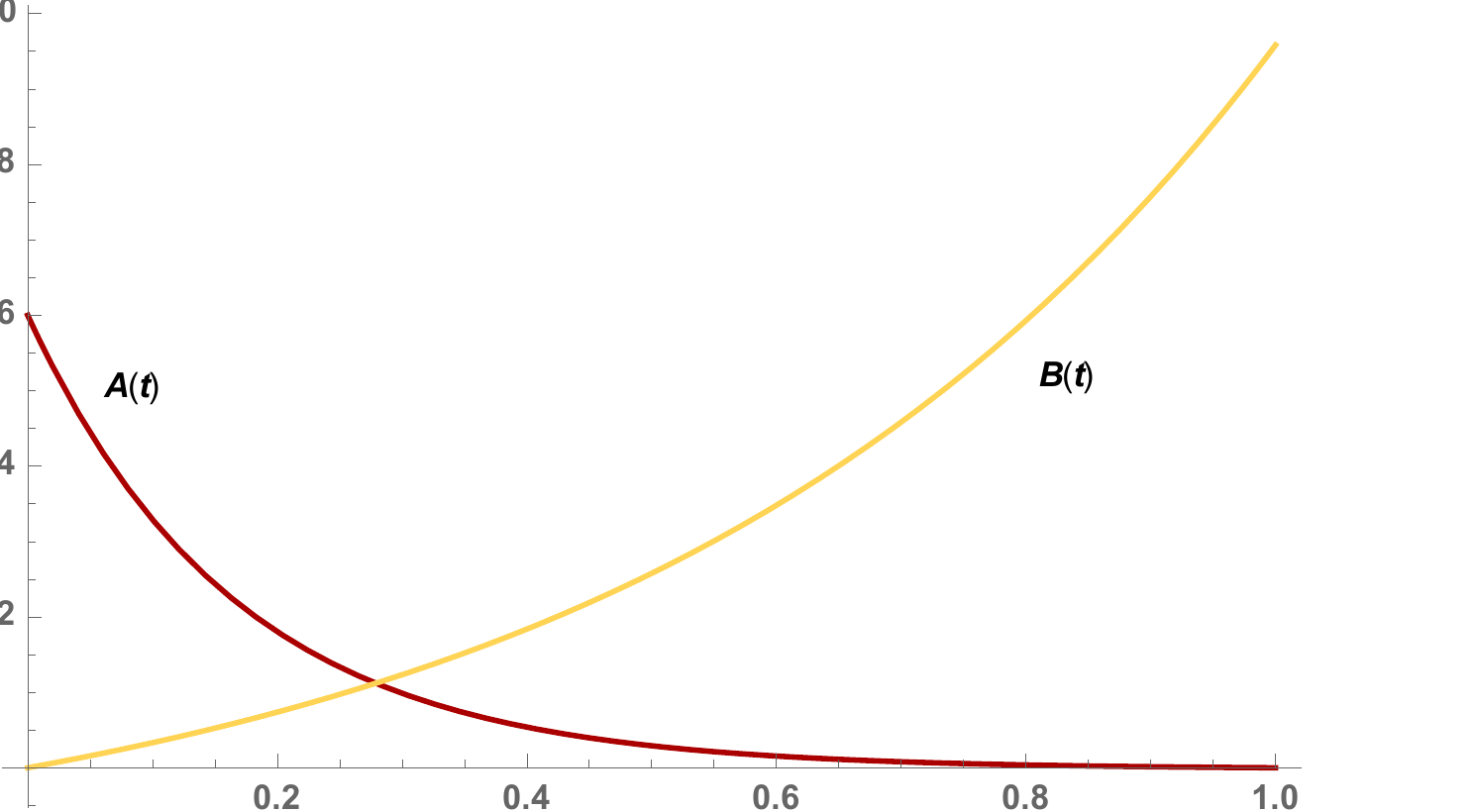}
	\caption{Plot of the smooth functions  $A(t)$ (red line) and $B(t)$ (yellow line)   defining the annealing schedule.}
	\label{schedule}
\end{figure}

The annealing process is scheduled as follows: at the beginning, the transverse field strength $A(t)$ is large i.e. $A(0) \gg B(0)$  and the  evolution is governed  by the $ H_0 $, responsible for quantum tunnelling effects;  $A(t)$ and $B(t)$ vary in time according to the graph in Fig.\ref{schedule} until, at the end of the annealing, $A(T_{\rm QA}) \ll B(T_{\rm QA})$ and the dominant   term in the evolution is  the problem hamiltonian $ H_P $.
\\

\FloatBarrier

\subsubsection{D-Wave Quantum Annealer}
The Canadian company \textit{D-Wave Systems} is nowadays the world leading producer  of quantum annealing devices  which  provides  access to their machines via cloud. The current most advanced QA  hardware  is  a 2048-qubit QPU going under the name of \textit{D-Wave 2000Q}.

D-Wave devices solve specific  combinatorial optimization problems  known as  Quadratic Unconstrained Binary Optimization (QUBO) problems, to which is associated an objective function of the  form
\begin{equation}
O(x)=\sum_{i}h_i {x}_i+ \sum_{i>j}J_{i,j} x_i x_j
\end{equation}
with $ x_i \in \left\lbrace 0,1 \right\rbrace  $  binary variables and  $ h_i $ and  $ J_{ij} $ parameters whose values encode the optimization task to solve.

The D-Wave  QA  hamiltonian  $ \mathcal{H}_{QA} $ is given by 
\begin{equation}
\begin{split}
\mathcal{H}_{QA}={A(s)}\sum_{i}\hat{\sigma}^{(i)}_x + {B(s)}H_P
\end{split}
\end{equation}
where annealing parameters $ A(s) $ and $ B(s) $ are  those  shown  in Fig.\ref{schedule}.
The problem Hamiltonian $ H_P $ associated to the objective function $ O(x) $ is expressed   as an  Ising hamiltonian 
\begin{equation}
\begin{split}
H_P=\sum_{i}h_i\hat{\sigma}^{(i)}_z + \sum_{i>j}J_{i,j}\hat{\sigma}^{(i)}_z \hat{\sigma}^{(j)}_z
\label{formula 1}
\end{split}
\end{equation}
where $ \hat{\sigma}^{(i)}_x $ and $ \hat{\sigma}^{(i)}_z $ are Pauli $ x $ and $ z $  operators.
Being $ \sigma^{(i)}_z \in \{+1,-1\} $, it is possible to switch between Ising and  QUBO  problems using the relation $ x_i=\frac{\sigma^{(i)}_z+1}{2} $. Hence the two formulations are perfectly equivalent, in the sense that  finding the ground state energy of the Ising problem corresponds to  solving the  associated QUBO.

Several interesting problems can be cast into the QUBO form,  examples include clustering \cite{kurihara2014quantum,Bottarelli2018}, graph partitioning and isomorphism \cite{Lucas}, map colouring \cite{Titiloye}, matrix factorization \cite{ottaviani2018low} but also portfolio optimization  \cite{Venturelli} and protein folding \cite{Perdomo}.

However, a crucial aspect to take into account when using the D-Wave annealer is the embedding  of  the \textit{logical} Ising/QUBO problem  into the physical  architecture of the quantum device, that we  address as  \textit{physical} Ising problem.

This mapping is usually found via  a heuristic algorithm named \textit{find\_embedding}, available through the D-Wave python libraries, which searches for the  minor-embedding of the  graph representing our  problem into the target hardware graph \cite{Cai2014}.
Once the mapping has been found, the D-Wave solves the embedded physical Ising problem where logical variables of the QUBO are represented as chains of physical variables (qubits). For this reason the embedding always increases the number of variables required in the optimization.

\section{Qubo Formulation of the prime factorization problem}\label{QUBO}
Consider   an integer  $N$ that could be represented in binary form using   $  L_N $   bits, 
\begin{equation}
\ceil{log_2(N)}= L_N
\end{equation} where $\ceil{\cdot}$ is the ceiling function. The  PF problem asks   to find  the  two prime factors $p$ and $q$ such that
\begin{equation}
N=p \times q.
\end{equation}
Expressing the above numbers $ N $, $ p $ and $  q $  in binary,    the following relations are obtained
\begin{equation}
N = \sum_{i=0}^{L_n-1}2^i n_i, ~
\label{N_binary}
\end{equation}
\begin{equation}
p = \sum_{j=0}^{L_p-1}2^j p_j
\ \ \ \ \ \ \ \mathrm{and} \ \ \ \ \ \ \ 
q = \sum_{k=0}^{L_q-1}2^k q_k.
\label{pq_binary}
\end{equation}
where $L_p$ and $L_q$ are respectively the bit-lengths of $ p $ and $ q $, while $n_i$, $p_j$ and $ q_k \in \{0,1\}$.
Since it can be proved that  either 
\begin{equation*}
L_n =L_p+L_q \hspace{1cm } \mathrm{or}  \hspace{1cm } L_n =L_p+L_q-1,
\end{equation*}
without loss of generality, the primes factors  $p$ and $q$ can be  chosen to be of the same length and specifically such that
\begin{equation}
L_p=L_q=\ceil{L_n/2}.
\label{length}
\end{equation}

Several approaches can be followed in order to convert problems like PF into an optimization problem.
Moreover, it often happens that the objective function $ O(p,q) $ involves terms with order higher than two. This means that $ O(p,q) $  is not in the desired QUBO  form but rather it is expressed as  a High-order Unconstrained Binary Optimization (HUBO) problem.

The trick used to transform HUBOs into QUBOs is to add ancillary  variables and substitute  high order terms with an expression having the same minimum and containing only quadratic terms \cite{Jiang}. As an example,    cubic term like $ \pm x_1\cdot x_2\cdot x_3 $ can be divided using the ancillary binary variable $a_1$ as 
\begin{equation}
\pm x_1\cdot x_2\cdot x_3\ \rightarrow\  \pm a_1x_3+2(x_1x_2-2x_1a_1-2x_2a_1+3a_1)
\label{trick}
\end{equation}
In the following, the main approaches used to convert  the factorization problem into HUBOs are presented.

\subsection{Direct Method}
A straightforward way   to  express PF  as an optimization problem  requires to consider the following objective function to minimize
\begin{equation}
O(p,q) = (N-p \cdot q)^2 
\label{objective}
\end{equation} where the square at the exponent  ensures that the  function has its global minimum only when $p$ and $q$ are the factors of $ N $. It is possible to rewrite Eq.\ref{objective}  in binary form  using Eq.s~\ref{N_binary}-\ref{pq_binary} as 
\begin{equation}
O(p,q) =  \left[ \left( \sum_{i=0}^{L_n-1}2^i n_i \right) -\left(  \sum_{j=0}^{L_p-1}2^j p_j\right)\cdot \left( \sum_{k=0}^{L_q-1}2^k q_k \right) \right]^2 
\label{objective2}
\end{equation}
where $ n_i $ are known coefficients, while the $ p_j $s and the $ q_k $s constitute the binary unknown variables among which the objective function is minimized.

Unfortunately, using  this method for factorizing larger numbers is hard because the square of $  p \cdot q $  in the objective function produces  many high order terms and a large number of ancillary variables are required \cite{Jiang}.

\subsection{Multiplication Table Method}
An alternative strategy \cite{Schaller} to translate   PF  into an  optimization problem involves the construction of  the  binary multiplication table of $ p $ and $ q $.
Without loss of generality,  we can write $ p $ and $ q $ in binary form   using   $ p_i $ and $ q_i \in \left\lbrace 0,1 \right\rbrace $ as
 \begin{equation}\begin{split}
p=(1,p_{L_p-2},...,p_3,p_2,p_1, 1) \\
q=(1,q_{L_q-2},...,q_3,q_2,q_1, 1)
 \end{split}
\end{equation}
where  $p_0 $ and $q_0  $ are set to $ 1 $ because $ p $ and $ q $   must be  odd. Also the most significant digits $\ p_{L_p-1}=q_{L_q-1}=1$ to ensure $ p $ and $ q $ have the correct length.

 Suppose   $N$  to be a number with binary length $L_N=8$, since, by Eq.\ref{length}, both  $ p $ and $ q $ must be  such that $L_p=L_q=4$, we can schematically represent Eq.s \ref{N_binary}-\ref{pq_binary} as
\begin{eqnarray}
%\hspace*{-0.3cm}
\begin{array}{|c |c|c|c|c|c|c|c|c|} 
\hline
&   2^7 & 2^6 & 2^5 & 2^4 & 2^3 & 2^2 & 2^1 & 2^0  \\ %[1.2ex]
% \hhline{~==========}
\hline
\hline
p&     &    &    &     & 1 & p_2 & p_1 & 1 \\ 
\hline
q&     &    &   &     & 1 & q_2 & q_1 & 1\\  
\hline 
N &  n_7 &  n_6 & n_5& n_4& n_3& n_2 & n_1 & n_0  \\
\hline
\end{array} \nonumber
\end{eqnarray}
\\
The binary multiplication $ p\cdot q  $ can be visualized using the following table
\begin{eqnarray}
%\hspace*{-0.3cm}
  \begin{array}{|c | c|c|c|c|c|c|c|c|} 
\hline
\multirow{1}{*}{ \footnotesize \textbf{Columns} } &   7 & 6 & 5 & 4 & 3 & 2 & 1 & 0  \\
\hline\hline
\multirow{4}{*}{ \footnotesize $ \mathbf{ p \cdot q }$ }&       &   &   &   & 1 & p_2 & p_1 & 1\\ \cline{2-9}
&     &    &    & q_1 & p_2q_1 & p_1q_1 & q_1 &   \\  \cline{2-9}
&     &   & q_2 & p_2q_2 & p_1q_2 & q_2 &     &   \\   \cline{2-9}
&    & 1 & p_2 & p_1 & 1 &     &   &\\   \cline{2-9}

\hline\hline
\multirow{2}{*}{\footnotesize \textbf{Carries}} & c_{67} & c_{56} & c_{45} & c_{34} & c_{23} & c_{12} & & \\  
 & c_{57} & c_{46} & c_{35} & c_{24} & & & & \\[2pt]

%\cline{2-11}
\hline \hline\hline
\footnotesize \textbf{N}  & n_7 &  n_6 & n_5& n_4& n_3& n_2 & n_1 & n_0  \\
\hline
\end{array} \nonumber
\label{mult_table}
\end{eqnarray}
where the bitwise multiplication is performed in columns. The $ c_{i,j} $ are the carries relative  to column $ j $ which came from  the sum  obtained in column $ i $. Carries are calculated assuming that all bits in each column are ones.

At this stage it is possible to construct a system of equations equating the sum of each  column $ i $, including its carries,  to the corresponding $ n_i $ shown in the last row.
In the previous example, the system of equations becomes
\begin{equation}
\begin{cases} p_1+q_1-2c_{12} = n_1 \\
p_2+p_1q_1+q_2+c_{12}-(2c_{23}+4c_{24})= n_2 \\
1+p_2q_1+p_1q_2+1+c_{23}-(2c_{34}+4c_{35})=n_3 \\
q_1+p_2q_2+p_1+c_{24}+c_{34}-(2c_{45}+4c_{46})=n_4\\
p_2+q_2+c_{45}+c_{35}-(2c_{56}+4c_{57})=n_5 \\
1+c_{56}+c_{46}-2c_{67}=n_6 \\
c_{57}+c_{67}=n_7
 \end{cases}
 \label{eq_sys}
\end{equation}

The objective function is obtained considering the square of each equation in the system
\begin{equation}
\begin{split}
& O(p,q) =  \left( p_1+q_1-n_1-2c_{12}\right)^2+\\ &+ \left( p_2+p_1q_1+q_2+c_{12}-n_2-2c_{23}-4c_{24}\right)^2 +\ ...\ +\\ &+\left( 1+c_{56}+c_{46}-n_6-2c_{67} \right) + \left( c_{57}+c_{67}-n_7\right)^2
\end{split}
\end{equation}
The HUBO obtained in this way usually contains less high order terms with respect to the Direct Method, 
 however it is possible to further improve this strategy, as shown in the following section.

\subsection{Blocks Multiplication Table Method}
This variation of the multiplication table method  is able to reduce the number of binary variables appearing in the optimization function by performing $ p \cdot q $ in blocks rather than in columns \cite{Jiang}. An example is shown in the table below, for the case of $L_p=L_q=4$.
\begin{eqnarray}
%\hspace*{-0.3cm}
\begin{array}{|c ? c|c|c? c|c?c|c?c |} 
\hline
\multirow{1}{*}{ \footnotesize \textbf{Blocks} } & \multicolumn{3}{c?}{\textbf{III}} & \multicolumn{2}{c?}{\textbf{II}}& \multicolumn{2}{c?}{\textbf{I} }& \multicolumn{1}{c|}{\textbf{} } \\
\hline\hline
\multirow{4}{*}{ \footnotesize $ \mathbf{ p \cdot q }$ }&       &   &   &   & 1 & p_2 & p_1 & 1\\ \cline{2-9}
&     &    &    & q_1 & p_2q_1 & p_1q_1 & q_1 &   \\  \cline{2-9}
&     &   & q_2 & p_2q_2 & p_1q_2 & q_2 &     &   \\   \cline{2-9}
&    & 1 & p_2 & p_1 & 1 &     &   &\\   \cline{2-9}

\hline\hline
\multirow{1}{*}{\footnotesize \textbf{Carries}} &  & c_{4} & c_{3} & c_{2} & c_{1} & & &\\

%\cline{2-11}
\hline \hline\hline
\footnotesize \textbf{N}  & n_7 &  n_6 & n_5& n_4& n_3& n_2 & n_1 & n_0  \\
\hline
\end{array} \nonumber
\label{blocks_mult_table}
\end{eqnarray}
\\
Bold vertical lines indicates that the table is split into three non trivial blocks. The rightmost column only contains a $ 1 $ and it is not taken into consideration since $ N $ is odd and also $ n_0 $ must be a $ 1 $.
The carries again are found assuming all bits in the columns inside each block to be ones. Those  generated form block I are  $ c_1 $ and $ c_2 $ while those generated form block II are $ c_3 $ and $ c_4 $.

It is worth noticing that, in this example, the blocks multiplication table strategy allowed  to reduce the number of carries from the 10 obtained   in the previous section to just 4. This constitute a great reduction in the number of QUBO variables, especially when  the number $ N $ to factor becomes large.

In order to solve the problem it is necessary to sum the contributions of each block   and equate it to the corresponding bits of $ N $. The following system of equations is obtained for the example related the previous table.
\begin{equation}
\begin{cases} 
(p_1+q_1)+2(p_2+p_1 q_1 +q_2)-(8c_2 + 4c_1) = n_1+2n_2 \\\\
(1+p_2 q_1+p_1 q_2 +1+ c_1)+2(q_1+p_2 q_2 +p_1 + c_2)-\\-(8c_4 + 4c_3)= n_3+2n_4 
\\\\
(q_2+p_2+c_3)+2(1+c_4) =  n_5+2n_6+4n_7
\end{cases}
\label{eq_sys_2}
\end{equation}

The objective function is again  obtained considering the square of each equation in the system \ref{eq_sys_2}
\begin{equation}
\begin{split}
& O(p,q) =  \left[ (p_1+q_1)+2(p_2+p_1 q_1 +q_2)-(8c_2 + 4c_1) -\right.\\ &\left.  -  (n_1+2n_2)\right]^2+ \left[(1+p_2 q_1+p_1 q_2 +1+ c_1)+ \right.\\ &\left.  +2(q_1+p_2 q_2 +p_1 + c_2)-(8c_4 + 4c_3)- (n_3+2n_4)    \right]^2+\\ &+ \left[ (q_2+p_2+c_3)+2(1+c_4) - ( n_5+2n_6+4n_7)\right]^2
\end{split}
\label{obj_funtion}
\end{equation}

The HUBO shown  in  Eq.\ref{obj_funtion} can be cast in the required QUBO format applying  the trick of  Eq.\ref{trick}.
\\

When solving such problem  using the D-Wave quantum annealer it is necessary to keep in mind two aspects. On one side, the device is limited in the number of logical variables it can embed while, on the other side, the range of coefficients appearing in the QUBO shouldn't span over a range of values that is too broad. If one of these two conditions is not satisfied, the  performances of the quantum annealing device could be severely affected.

As a matter of fact, from Eq.\ref{obj_funtion}, it is clear that the range of  coefficients appearing in the optimization function strictly depends on the size of the blocks, i.e. on the number of columns contained in each block. 
Therefore, an appropriate choice of the blocks number and sizes should take into account for  the trade-off between number of QUBO variables in play  and range of the coefficients.
Approximatively, an acceptable splitting into blocks makes the largest  QUBO coefficient  no larger than $ O((log(N))^3) $ \cite{Jiang}.

\section{Prime Factorization on The low noise D-Wave 2000Q}
\label{PrimeDWave}
In this section, a quantum annealing approach to the problem  of PF is presented and discussed.

 Our aim is the study of the  performances and limits  of the \textit{low noise D-Wave 2000Q} in addressing such problem.
For this reason, we challenged the annealing device to perform a factorization of several integers $ N $, gradually increasing their size, i.e. the bit-length $ L_{N} $. 
The numbers selected for this study are shown in the table below.

\begin{table}[htbp]
	\begin{center}
		\begin{tabular}{ | p{1.5cm} |p{1.5cm} |p{1.5cm} |p{1.7cm} |}
			\hline
		\centering	 \textbf{\textit{N}}  & \centering \textbf{\textit{p}}  & \centering \textbf{\textit{q}}   &  \textbf{Length of  \textit{N}  in bits, $L_N$}   \\ \hline
			143 &  13 & 11 & 8   \\ \hline
			3127 &  59 & 53  &12   \\ \hline
			8881	 &  107 & 83  & 14\\ \hline
			59989	 &  251 & 239  & 16\\ \hline
			103459	 &  337 & 307  & 17\\ \hline
			231037	 &  499 & 363  & 18\\ \hline
			376289	 &  659 & 571  & 19\\ \hline
		\end{tabular}	
		\caption{Selected numbers $ N $ to factor, associated primes $ p $ and $ q $  and bit-length of $ N $.}
		\label{tab_N}		
	\end{center}

\end{table}\FloatBarrier 
For each   $  N $ in Table \ref{tab_N} we applied the Block Multiplication Table method for finding its prime factors. The HUBO optimization function obtained is then transformed to QUBO via the      \textit{make\_quadratic} function available through D-Wave \textit{dimod} libraries \cite{DW}.
In the following, we report the properties of the QUBOs graph structure obtained  varying $ N $.

\subsection{Number of logical variables and connectivity}
\begin{figure}[htbp]
	\includegraphics[scale=0.55]{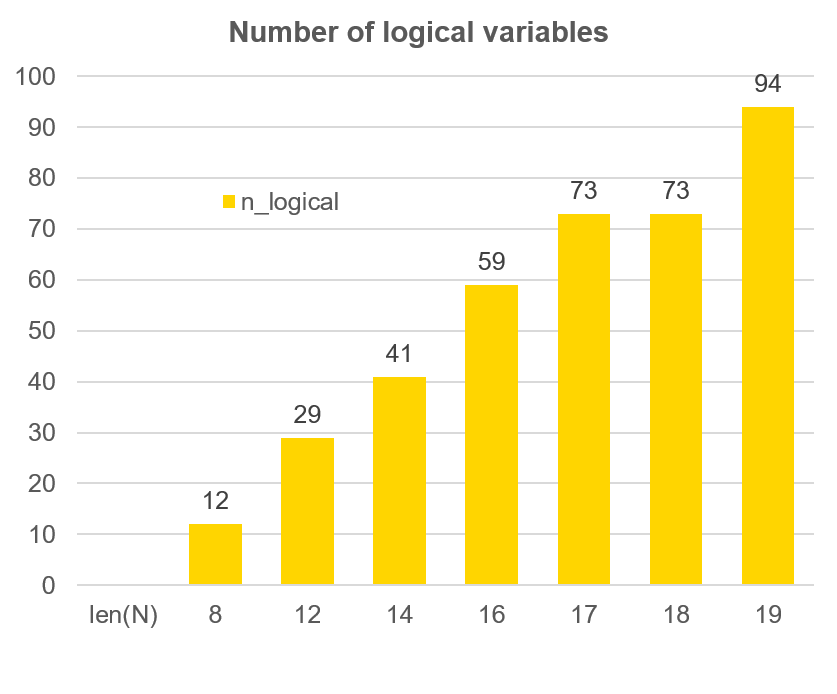}				
	\caption{Number of  logical variables appearing in the QUBO in relation to an increasing bit-length of  $ N $. }	
	\label{n_logical}
\end{figure}

\begin{figure}[htbp]
	\includegraphics[scale=0.55]{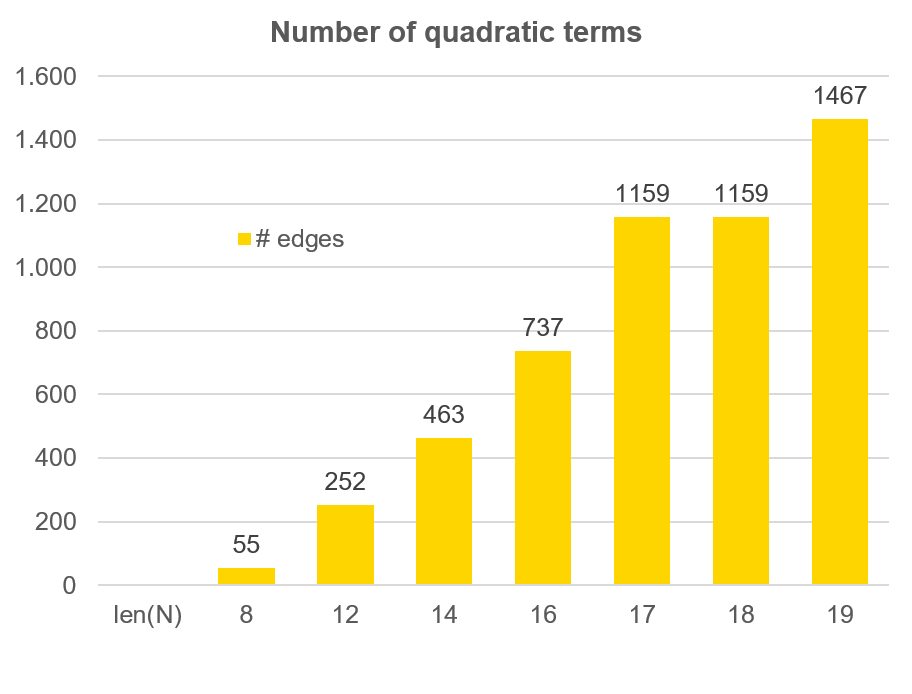}		
	\caption{ Number of quadratic terms (edges) appearing in the QUBO in relation to an increasing bit-length of  $ N $. }	
		\label{edge_logical}
\end{figure}

The number of QUBO logical variables increase polynomially with respect to $ L_N $, as shown in Fig. \ref{n_logical}. 
Also the QUBO quadratic terms  seem to grow polynomially with the problem size (see Fig. \ref{edge_logical})
\FloatBarrier 
\FloatBarrier

\begin{figure}[htbp]
	\includegraphics[scale=0.4]{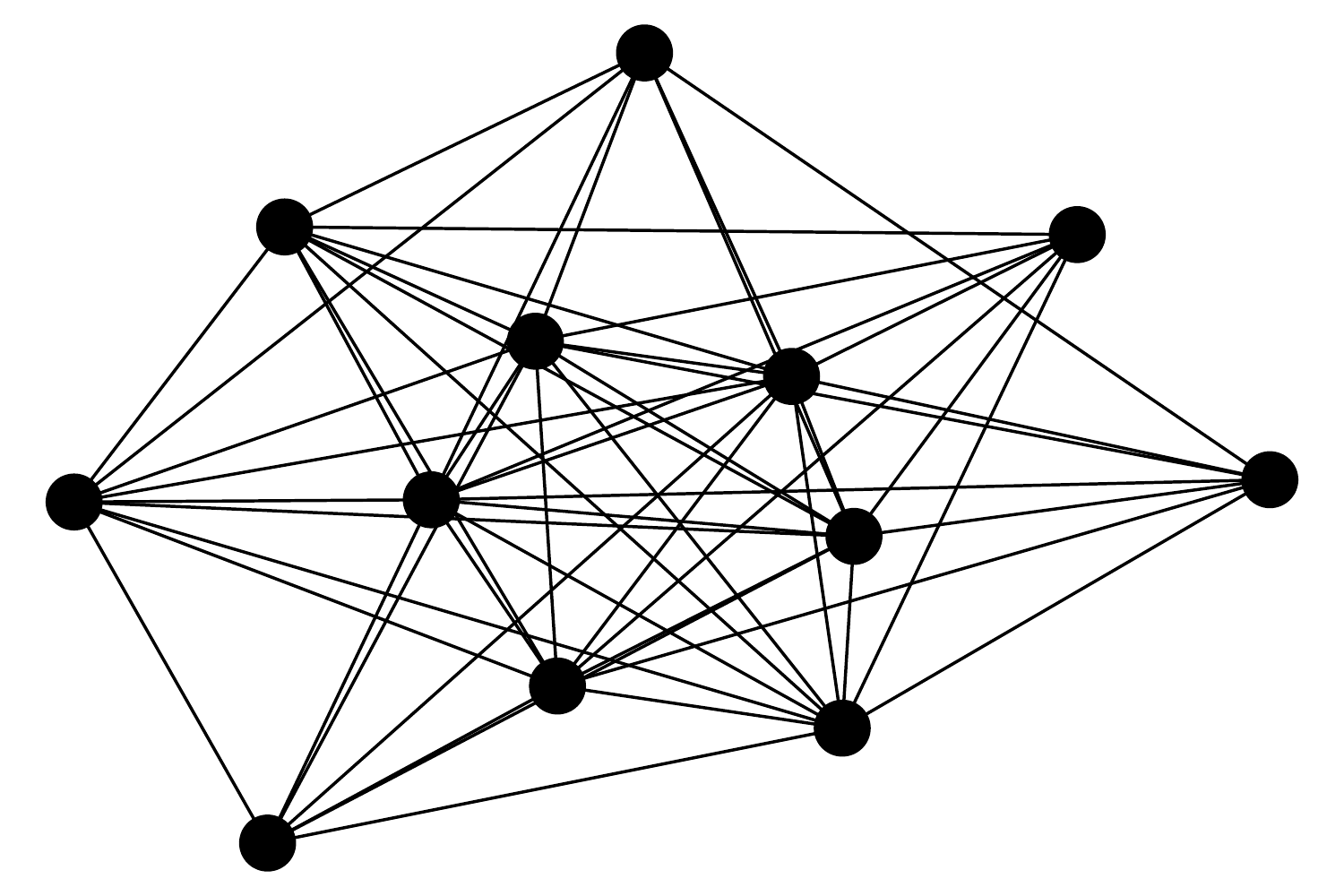}
	\caption{Graph structure of the QUBO problem related to the factorization of 143. The QUBO involves 12 binary variables (nodes) and 55 quadratic terms (edges). }
	\label{143_connectivity}
\end{figure}

In the graphical representation of the QUBO, quadratic terms identifies edges between the logical variables, which constitute the nodes of the graph. An example is shown  in Fig.\ref{143_connectivity} where it is graphically represented the  structure of the QUBO associated to the factorization of $ N=143 $.
\FloatBarrier \noindent
The QUBO for $ N=143 $ involves 12 binary variables:
\begin{itemize}
	\item[-] $ p_1, p_2, q_1$ and $ q_2 $ are those related to the factors $ p $ and $ q $;
		\item[-] $ c_1, c_2, c_3$ and $ c_4 $ are the carries;
		\item[-] $ a_1, a_2, a_3$ and $ a_4 $ are the ancillary variables needed to split high order terms into quadratic.
\end{itemize}

\subsection{Embedding}
\begin{figure}[htbp]
	\includegraphics[scale=0.5]{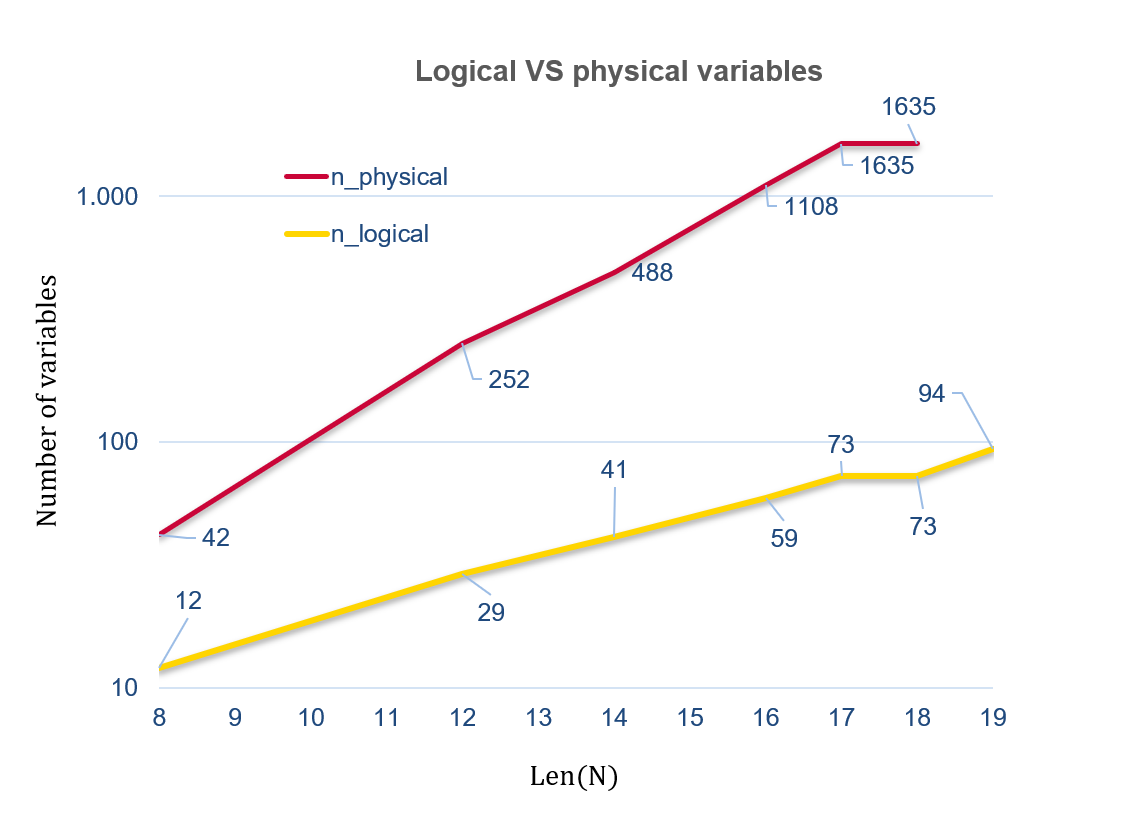}		\vspace{-.5cm}
	\caption{Comparison (in log scale) between number of logical variables in the QUBO and median number of physical variables (qubits) needed in the associated embedded Ising problem.}
	\label{embedding}
\end{figure}

In order to use the D-Wave for the PF problem, it is necessary to embed the  logical QUBOs  into the actual topology of the machine. This mapping is performed via a heuristic minor-embedding approach. In Fig.\ref{embedding},  the comparison
between logical variables and the median number of physical qubits required after the embedding is reported.
\begin{figure}[htbp]
	\includegraphics[scale=0.55]{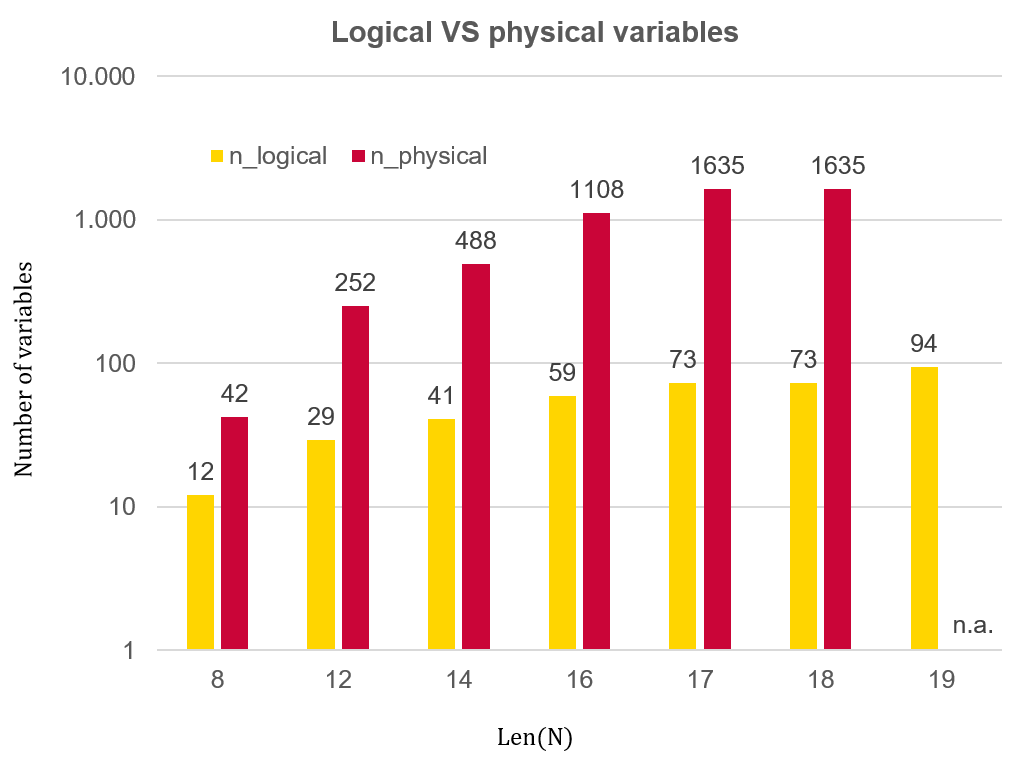}	
\end{figure}
\FloatBarrier
\FloatBarrier 
The D-Wave algorithm \textit{find\_embedding} was able to heuristically find an embedding for all instances, except for  $ N=376289 $, i.e. $ L_N=19 $. This is due to the fact that the problem is approaching the annealer limit in the number of physical variables (qubits) required. Actually it  is possible find an \textit{ad hoc} manual embedding for this instance as claimed in \cite{Jiang}, however, this goes beyond the purpose of our work since our aim  is to test the D-Wave device both from the hardware and software perspective.

\begin{figure}[htbp]
	\includegraphics[scale=0.45]{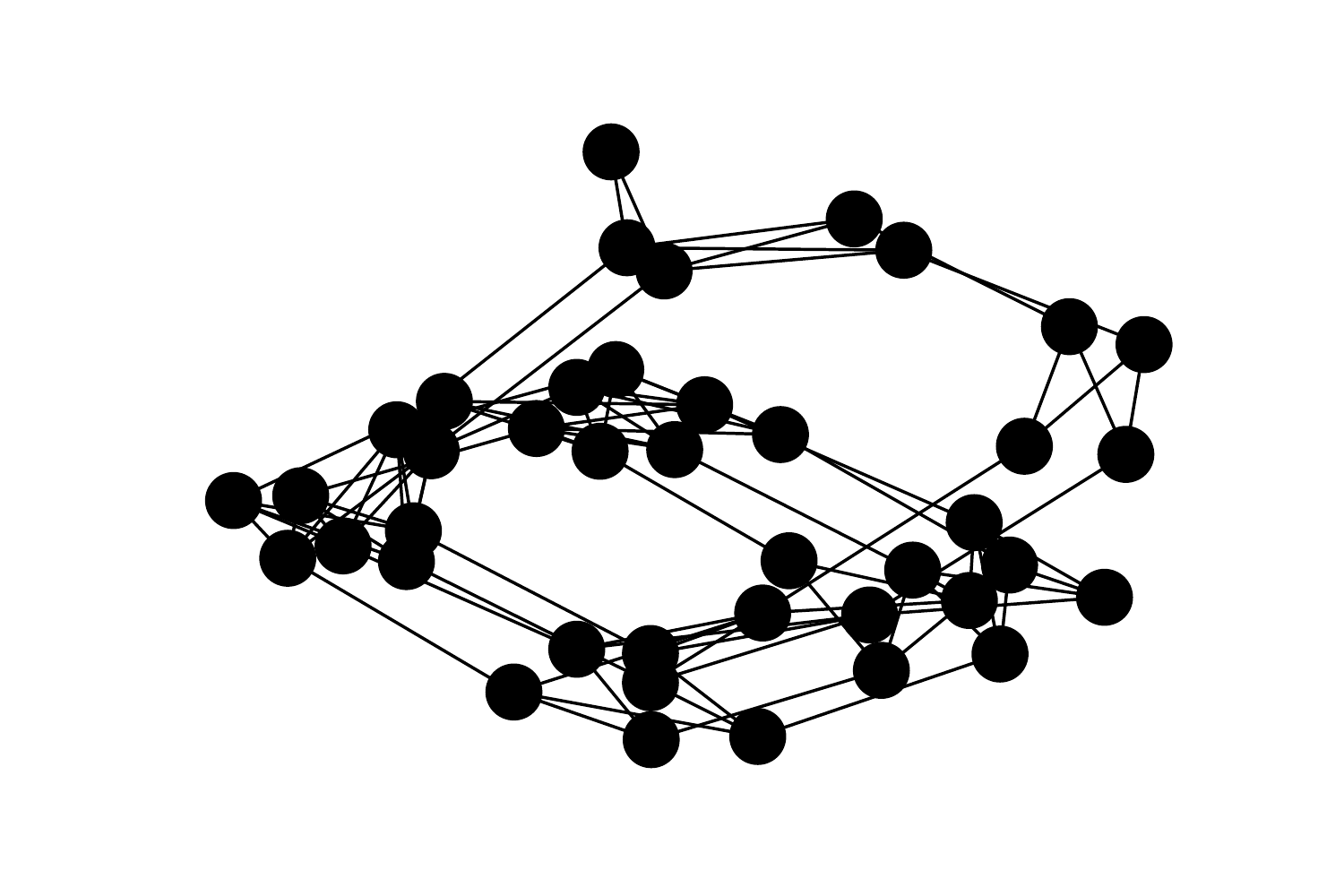}		\vspace{-.5cm}
	\caption{Graph structure of the embedded Ising problem related to the factorization of 143. The embedded problem involves 42 physical qubits (nodes) and 93 coupling terms (edges).  }
\label{143_connectivity_embedded}
\end{figure}

Finally,  form the trends shown in Fig.\ref{embedding}, it is clear that the gap between the number of logical variables in the QUBO and the physical qubits in the embedded Ising rapidly increases with the problem size. This currently constitute one of the main limitations of the  D-Wave device which could be overcome with a more connected topology of the device.
As an example, Fig.\ref{143_connectivity_embedded} shows the graph of  the embedded Ising problem associated to the factorization of $ N=143 $.
The difference in complexity with respect to the logical QUBO for the same problem, shown in Fig.\ref{143_connectivity}, is remarkable.
\FloatBarrier

\subsection{Results}
\begin{figure}[htbp]
	\includegraphics[scale=0.5]{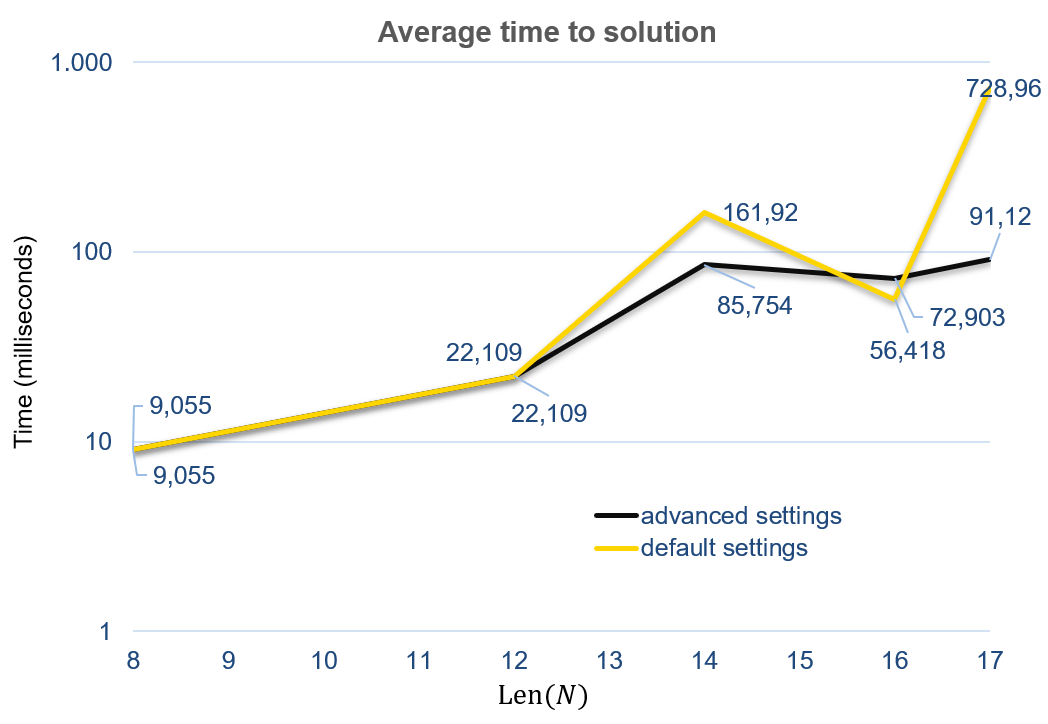}		
	\caption{ Time To Solution (TTS) needed for the D-Wave to output a correct solution to the factorization problem as a function of the bit-length of $ N $. Solutions found using default annealing setting are identified by the   yellow line; the black line shows the TTS found using advanced settings. }
	\label{result}
\end{figure}

The metric used to evaluate the low noise D-Wave 2000Q device was  Time To Solution (TTS) i.e.  the average time, expressed in milliseconds, needed by the quantum device to find the prime factors $ p $ and $  q $. Such quantity is defined as follows
\begin{equation}
\mathrm{TTS}= \dfrac{total\  QPU\ access\ time}{number \ of\ times\  N\ is\ factored\ correctly}
\end{equation}
where  the total  QPU access time  is defined as the sum of the  programming and sampling time required by the quantum annealer \cite{QPUtime}. 
D-Wave runs have been performed in a regime of forward annealing, setting the annealing time for a single anneal to $ 1 \mu s $ and $ 10000 $  number of reads.
Results of the computational times are reported in the graph shown in Fig.\ref{result}. 
\FloatBarrier 
As a  first approach, default settings of the D-Wave have been used (yellow line in Fig.\ref{result}). Subsequently, advanced techniques have been used to increase the probability of solving the optimization problem.
The advanced settings  we selected are:
\begin{itemize}
	\item\textit{ extended J range}, which enables chains of physical qubits representing a single logical variable to be coupled with an higher strength $ J_{chain} = -2 $ keeping all other non-chain couplings in the range $ \left[ -1,1 \right] $;
	\item \textit{ annealing offset}, which delays the evolution of qubits applying offset in the annealing paths of the qubits, so that some are annealed slightly before others, improving performances. 
\end{itemize}
The main improvement was obtained in the largest instance solved $ N=103459 $, i.e. $ L_N=17 $, where the  TTS is decreased by almost an order of magnitude (from $ 728.96 \mu s $ to $ 91.12 \mu s $).

\section{Conclusions and Future Work}
RSA  \cite{RSA} is a widely used cryptographic algorithm for securing data based on a fundamental  number theory problem known as Prime Factorization (PF).
The hardness of such problem makes harmless any attack performed by a classical computer since the computational time needed to factor a  number 
$ N  $ grows exponentially with the size of $ N $.

A  quantum computer able to run Shor's algorithm instead could factor the same number $ N $ only in polynomial time with respect to its size \cite{Shor}.
Such advantage, despite being theoretically solid, can be achieved only by a large scale fault tolerant general purpose quantum computer, which is  far from being built in the near term.

In recent years, a large effort has been put forward in the development of annealers,  with  different   Shor-like quantum annealing algorithms proposed in the literature \cite{Schaller,Dattani,Jiang}.

For this reason, with the present work, we aimed at analysing three different % approaches to PF using the most advanced quantum annealer available, i.e. the \textit{low noise D-Wave 2000Q}.
 methods for addressing the  PF problem with a quantum annealer. Among them, we selected the most advanced  (\textit{block multiplication table method}) to test the  performances and limitations  of  the \textit{low noise D-Wave 2000Q} quantum annealer.
The   device was challenged to perform a factorization of several integers $ N $, gradually increasing their size, going from $ L_N=8 $ to $ 19 $.

The heuristic algorithm \cite{Cai2014} needed to map PF into the D-Wave architecture successfully embedded problems up to $ L_N=18 $. Advanced techniques like extended J range and annealing offsets have been employed in the quantum annealing optimization in order to reach superior performances. The \textit{low noise D-Wave 2000Q}  factored correctly all integers $ N $ up to $ L_N=17 $, with a time to solution (TTS) that never exceeded the $ 100 $ milliseconds in  the advanced setting.

\begin{figure}[htbp]
	\includegraphics[scale=0.55]{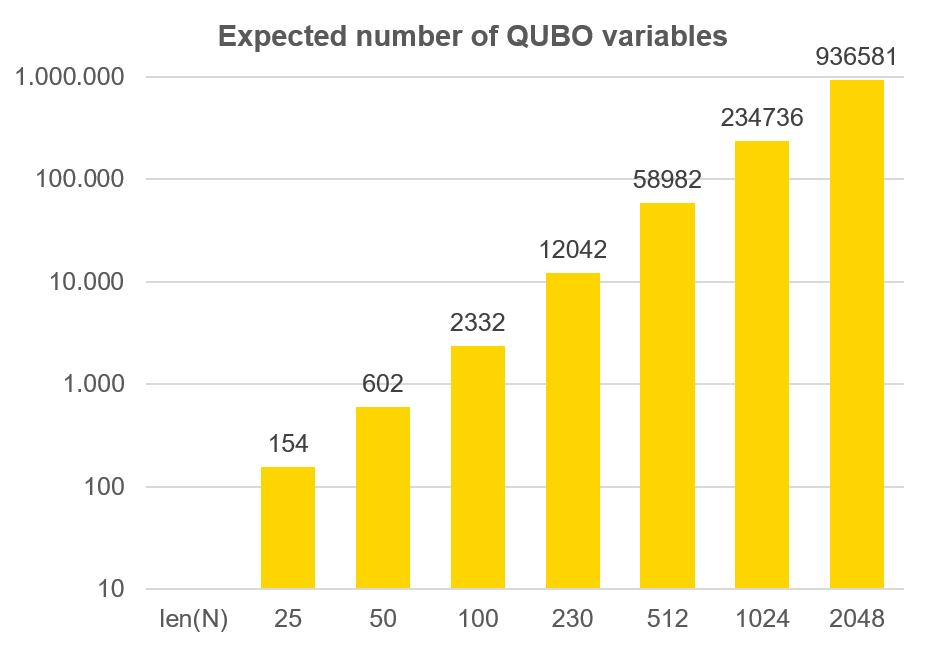}		
	\caption{Expected trend of the number of QUBO logical variables needed for the factorization problem as the bit-length of $ N $ increases.}
		\label{expected_n}
\end{figure}

The obtained results are promising, however further progresses on the quantum annealing hardware  are required to factor larger numbers. 
The figure below  (Fig.\ref{expected_n}) shows the expected number of logical variables required by the QUBO to solve PF increasing the problem size.
\FloatBarrier
It is clear that hardware improvements should focus  on  increasing the qubit count and   connections in order to solve problems of such complexity. 
For this reason, it will be the subject of  future studies the test of  the factoring performances of the new generation D-Wave device named \textit{Advantage} as well as the employment of hybrid quantum-classical techniques \cite{Peng} able to divide large QUBOs into sub-problems, individually solved by quantum annealing.

%Finally, in light of the increasing interest in the development of hybrid quantum-classical approaches,  it will be our 

\end{document}